\documentclass[aps,prappl,reprint,showpacs]{revtex4-1}

\usepackage{amsmath}
\usepackage{amssymb}
\usepackage{graphicx}
\usepackage{hyperref}

\renewcommand{\Re}{{\rm Re}}
\begin{document}

\title{ Two-dimensional optomechanics formed by the  graphene sheet and photonic crystal cavity}

\author{Hui Wang}
\author{Qifeng Qiao}
\author{Chenyu Peng}
\author{Ji Xia}
\author{Guangya Zhou}\email{mpezgy@nus.edu.sg}
\affiliation{Department of Mechanical Engineering, National University of Singapore, Singapore 117576}

\author{Yan-Jun Zhao}
\affiliation{Beijing National Laboratory for Condensed Matter Physics,
Institute of Physics, Chinese Academy of Sciences, Beijing 100190, China\\
School of Physical Sciences, University of Chinese Academy of Sciences, Beijing 100190, China}

\author{Xun-Wei Xu}
\affiliation{Department of Applied Physics, East China Jiaotong University, Nanchang, 330013, China}

\date{\today}

\begin{abstract}
We theoretically study  photon transmission and mechanical ground state cooling in a two-dimensional optomechanical system that is formed by suspending a graphene sheet on a one-dimensional optomechanical crystal. When the frequencies of graphene resonator and nanobeam resonator(phononic mode of optomechanical crystal)  are approximately the same, the $\Lambda$-type degenerate four-level structure of two-dimensional optomechanical system  shows the two-color optomechanically-induced transparency, and the transparency window could be switched among probe signal's absorption, transparency, and amplification. According to our calculations, the graphene resonator could also  assist the ground state cooling of nanobeam resonator in the proposed two-dimensional optomechanics.

\pacs{42.79. Gn, 42.50.Wk, 42.50.Lc}
\end{abstract}

\maketitle \pagenumbering{arabic}

\section{Introduction}

Cavity optomechanics supports a platform to explore  nonlinear and nonclassical effects \cite{Vahala,Painter2,Thompson,Aspelmeyer1,Poot}. Significant progresses have been made on the strong optomechanical coupling and  mechanical ground state cooling
 in optomechancis \cite{Marquardt2,Wilson,Schliesser,Vanner,Chan,Hertzberg,Mathew,Lecocq,Cohadon,Park}.
 The photonic nonlinear and nonclassical effects induced by optomechanical interaction have been widely studied, such as the OMIT (optomechanically-induced transparency) \cite{Agarwal,Agarwal2,Agarwal3,Kippenberg,Painter,Luki},  photon blockade \cite{Rabl,xu1,Li,Liao1,Liao2,Kurn,Liew,Savona}, nonclassical mechanical motions \cite{Teufel,Alegre,Cohadon,Park,xu3,xu4}, optical nonreciprocity \cite{Dong,Peterson,Fang,Lipson}, and so on. The hybrid system formed by optomechanics coupling to TLS (two-level system) or mechanical resonator also attracts a lot of interest recently \cite{Bienert,Ma,Hui1,Hui2,Hui3,Akram}.

Cavity optomechanics has been realized in different systems, such as Whispering-gallery cavity, photonic crystal cavity, microwave circuit, dielectric membrane placed between two high-finesse mirrors, and so on \cite{Vahala,Painter2,Harris}.  Graphene sheet has been used to build optomechanics with microwave resonator through radiation pressure interaction \cite{Singh,Bachtold,Weber,Song}.
  Moreover, the interaction between photonic crystal cavity and graphene sheet are more interesting, because the photonic cavity might be able to detect mechanical motions and nonlinear mechanical properties of graphene sheet with extra high precision \cite{Eichler}. In addition recent experiments have shown that the graphene sheet could  be used to tune the optical, electrical, and heat transport of photonic crystal cavity \cite{Gan,Huan,Xu,Majumdar,Barton,Shih}.

  However, single layer graphene is almost transparent for visible and infrared lasers \cite{Falkovsky,Geim}, so it is difficult to build optomechanics with single layer graphene sheet through radiation pressure type interaction. As verified by the perturbation  calculations and  recent experiments \cite{Gan,Meade,Huan}, the graphene sheet leads to resonant frequency shift and an additional damping for  photonic cavity mode through optical gradient or absorption force. Thus, the optical gradient or absorption force could  be used to build 2D (two-dimensional) optomechanical system consisting of the graphene sheet and 1D (one-dimensional) optomechanical crystal; as for radiation pressure type optomechanical interaction, the  multi-layer graphene sheets might be a choice.

With a suspended single-layer graphene sheet above a 1D optomechanical crystal,  we study photon transmission and mechanical ground state cooling in 2D optomechanics (see Fig.\ref{fig1}). The optomechanical type interactions exist not only between photonic and  mechanical modes of nanobeam, but also between the photonic cavity field and graphene resonator.  Compared with other type multi-mode optomechanics, the steady position of graphene resonator could be easily tuned via a control voltage between graphene sheet and silicon substrate \cite{Bunch,Chen}; thus, the damping rate of photonic cavity and graphene-cavity optomechanical interaction strength could be controlled. Here we mainly focus on the  effects of a graphene sheet on photon transmission and ground state cooling of the nanobeam resonator in proposed 2D optomechanical system.

The paper is organized as follows: In Section II, we introduce the proposed 2D optomechanics model. In Section III, we analyze the effects of a graphene resonator on photon transmission. In Section IV, we study the graphene resonator assisted ground state cooling of nanobeam resonator. In Section V, we conclude this work.

\section{2D optomechanical model}

\begin{figure}
\includegraphics[bb=15 60 585 420, width=8 cm, clip]{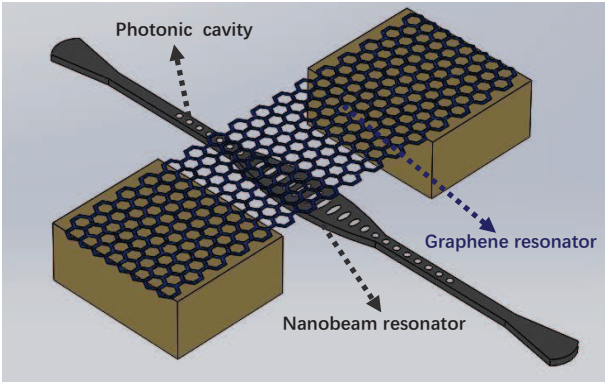}
\caption{(Color online) Two-dimensional Optomechanics formed by a suspended Graphene sheet on
an one-dimensional optomechanical crystal.}
\label{fig1}
\end{figure}

We study a 2D optomechanical system as shown in Fig.~\ref{fig1}, the photonic and phononic modes of nanobeam in an 1D optomechanical crystal interact with each other in a plane parallel to silicon substrate; the  single-layer graphene sheet interacts with photonic mode of 1D optomechanical crystal in the direction perpendicular to silicon substrate.  Thus, the Hamiltonian of free 2D optomechanical system:
\begin{eqnarray}\label{eq:1}
H_{0}&=&\hbar\omega_{a}\hat{a}^{\dagger}\hat{a}+\hbar\omega_{b}\hat{b}^{\dagger}\hat{b}+\hbar\omega_{c}\hat{c}^{\dagger}\hat{c}+\hbar g \hat{a}^{\dagger}\hat{a} (\hat{b}^{\dagger}+\hat{b})\nonumber\\
& &+\hbar \lambda \hat{a}^{\dagger}\hat{a}(\hat{c}^{\dagger}+\hat{c}).
\end{eqnarray}
Here $\hat{a}$ ($\hat{a}^{\dagger}$), $\hat{b}$ ($\hat{b}^{\dagger}$), and $\hat{c}$ ($\hat{c}^{\dagger}$) are annihilation (creation) operators of photon, nanobeam phonon, and graphene phonon, respectively.  $\omega_{a}$ is the resonant frequency of photonic cavity, and $\omega_{b}$ ($\omega_{c}$) corresponds to vibration frequency of nanobeam (graphene) resonator. And, $g$ and $\lambda$ describe the optomechanical interaction strengths of the photonic mode with the nanobeam and graphene resonators, respectively.
The graphene-cavity optomechanical coupling strength and cavity damping rate could be changed by adjusting graphene sheet's steady position which could be easily adjusted by a control voltage  between graphene and silicon substrate \cite{Bunch,Chen}.

\begin{figure}
\includegraphics[bb=10 320 590 770, width=8.0 cm, clip]{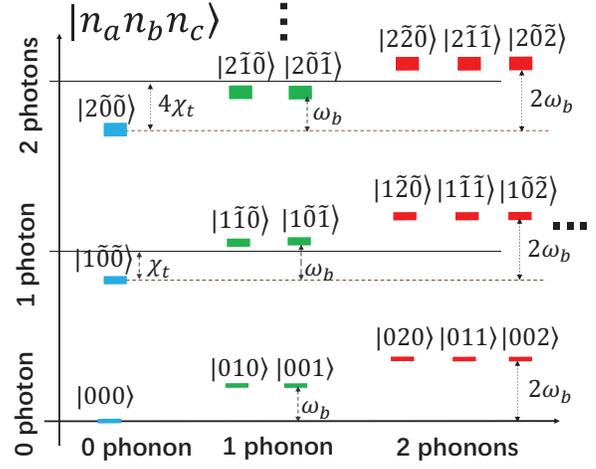}
\caption{(Color online) The energy level diagram of 2D degenerate optomechanical system ($\omega_{b}=\omega_{c}$).
Here $|n_{a}\rangle$, $|n_{b}\rangle$, and  $|n_{c}\rangle$ are the Fock states of photon and phonons, while
$|\tilde{n}_{b}\rangle$ and $|\tilde{n}_{c}\rangle$ correspond to phonon displaced Fock states.}
\label{fig2}
\end{figure}

In a rotating frame defined by an unitary transformation $U_{1}=\exp{\{\hat{a}^{\dagger}\hat{a} [g(\hat{b}^{\dagger}-\hat{b})/\omega_{b}+\lambda(\hat{c}^{\dagger}-\hat{c})/}{\omega_{c}}]\}$,  the effective Hamiltonian becomes $H_{eff}=\hbar(\omega_{a}-\chi_{t})\hat{a}^{\dagger}\hat{a}-\hbar\chi_{t} \hat{a}^{\dagger}\hat{a}^{\dagger}\hat{a}\hat{a}+\hbar\omega_{b}\hat{b}^{\dagger}\hat{b}+\hbar\omega_{c}\hat{c}^{\dagger}\hat{c}$.
The direct photon-phonon interactions  disappear in $H_{eff}$ and are replaced by a Kerr  nonlinearity term  with
the coefficient $\chi_{t}=\chi_{b}+\chi_{c}$, where $\chi_{b}=g^{2}/\omega_{b}$ and  $\chi_{c}=\lambda^{2}/\omega_{c}$
 are the photon nonlinear coefficients induced by nanobeam and graphene resonators, respectively.
 The energy level structure of 2D optomechanical system could be obtained from $H_{eff}$ as
\begin{eqnarray}\label{eq:2}
 E(n_{a},n_{b},n_{c})=\hbar\left(\omega_{a}n_{a}-\chi_{t}n^{2}_{a}\right)+\hbar\omega_{b}n_{b}+\hbar\omega_{c}n_{c}.
 \end{eqnarray}
 Here $n_{a}$ is the cavity photon number, while $n_{b}$ and $n_{c}$ are phonon numbers on nanobeam and graphene resonators, respectively.
 Compared with the energy level structure of the standard optomechanics\cite{Rabl,Aspelmeyer1,Girvin}, the energy levels in 2D optomechanics have an additional mechanical freedom. For degenerate mechanical modes $\omega_{b}=\omega_{c}$, the degree of energy level degeneracy could be calculated from Eq.~(\ref{eq:2}) as $n_{b}+n_{c}+1$.  The low excitation state energy levels are shown in Fig.~\ref{fig2}, the degree of degeneracy for zero phonon state is $1$, and it is $2$ (or $3$) for single (or double) phonon excited states, respectively. From the expression of $H_{eff}$, the  eigenstates of 2D optomechanical system can be obtained as $|n_{a}\tilde{n}_{b}\tilde{n}_{c}\rangle=U_{1}|n_{a}n_{b}n_{c}\rangle$, where $|n_{i}\rangle$ ($i=a,b,c$) are the Fock states of photon and phonons, respectively, while $|\tilde{n}_{b}(n_{a})\rangle$ and $|\tilde{n}_{c}(n_{a})\rangle$ are the phonon displaced Fock states in the case of photon number $n_{a}$.

\section{Tunable Photon transmission}

\subsection{Controllable Photons blockade}

The photon blockade in optomechanical system has been widely studied \cite{Rabl,xu1,Li,Liao1,Liao2,Kurn}.
 If the photonic cavity is driven by a weak laser field with frequency $\omega_{d}$ and amplitude $\Omega$, that is
$H_{d}=i\hbar [\Omega\exp{(-i\omega_{d}t)}\hat{a}^{\dagger}-h.c. ]$.
 Thus, the Hamiltonian in a rotating frame at driving frequency $\omega_{d}$, defined by an unitary transformation $U_{2}=\exp{(-i\omega_{d } \hat{a}^{\dagger}\hat{a} t )}$,  becomes
\begin{eqnarray}\label{eq:3}
H_{r}&=&-\hbar\Delta\hat{a}^{\dagger}\hat{a}+\hbar\omega_{b}\hat{b}^{\dagger}\hat{b}+\hbar\omega_{c}\hat{c}^{\dagger}\hat{c}+\hbar g \hat{a}^{\dagger}\hat{a} (\hat{b}^{\dagger}+\hat{b})\nonumber\\
& &+\hbar \lambda \hat{a}^{\dagger}\hat{a}(\hat{c}^{\dagger}+\hat{c})+i\hbar\left(\Omega \hat{a}^{\dagger}-\Omega^{\ast}\hat{a}\right).
\end{eqnarray}
here $\Delta=\omega_{d}-\omega_{a}$ is the frequencies detuning of the driving laser and photonic mode. Define $\kappa_{a}$, $\kappa_{b}$, and $\kappa_{c}$ as the
damping rates of photonic cavity, nanobeam resonator, and graphene resonator, respectively.
The photon  blockade  requires a weak driving field ($|\Omega|\ll\kappa_{a}$)  and it is usually described by the second-order correlation function which could be defined with density operator as $g^{(2)}(0)=\text{Tr}(\rho \hat{a}^{\dagger 2}\hat{a}^{2})/\text{Tr}(\rho \hat{a}^{\dagger}\hat{a})^2$.
The master equation of density operator:
\begin{eqnarray}\label{eq:4}
\dot{\rho}=\frac{1}{i\hbar}[H_{r},\rho]+L_{a}(\rho)+L_{b}(\rho)+L_{c}(\rho).
\end{eqnarray}
 The Lindblad terms in Eq.~(\ref{eq:4}) are: $L_{o}(\rho)=\frac{\kappa_{o}}{2}(n^{T}_{o}+1)(2\hat{o}\rho \hat{o}^{\dagger}-\hat{o}^{\dagger}\hat{o}\rho-\rho \hat{o}^{\dagger}\hat{o})+\frac{\kappa_{o}}{2}n^{T}_{o}(2\hat{o}^{\dagger}\rho \hat{o}-\hat{o}\hat{o}^{\dagger}\rho-\rho \hat{o}\hat{o}^{\dagger})$, with $o=a,b,c$ corresponding to optical and mechanical variables, respectively. Under weak pumping, the density matrix $\rho$ can be numerically calculated by truncating to limited photon and phonon numbers in  Eq.~(\ref{eq:4}) \cite{Liew,Tan,Johansson1,Johansson2}. The thermal photon and phonon numbers are defined as $n^{T}_{o}=[{\exp{[\hbar\omega_{o}/(k_{B}T)]}-1}]^{-1}$ ($o=a,b,c$), with $T$ and  $k_{B}$ are environmental temperature and Boltzmann constants, respectively. Because of the extra high frequency,  in this article the thermal photon numbers can be set as zero, that is $n^{T}_{a}=0$.

\begin{figure}
\includegraphics[bb=15 15 585 520, width=8.3 cm, clip]{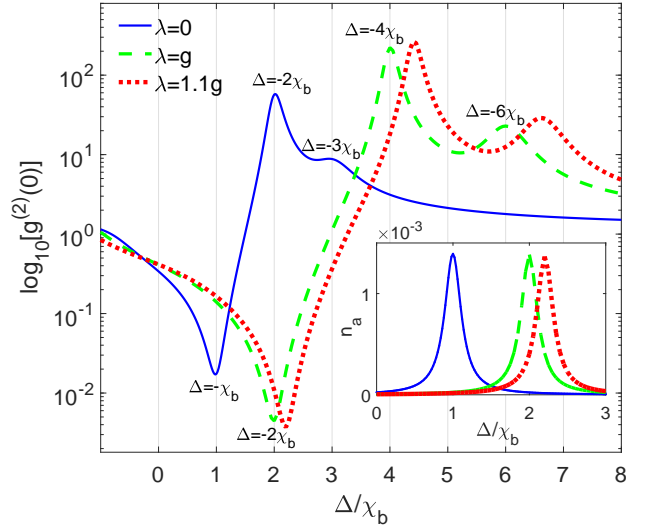}
\caption{(Color online) Logarithmic plot of  equal-time
 second-order correlation function $g^{(2)}(0)$ as the function of detuning $\Delta=\omega_{d}-\omega_{a}$,
with $\omega_{b}=\omega_{c}$ and $\chi_{b}=g^{2}/\omega_{b}$.
The blue-solid curve describes photon blockade of one-dimensional optomechanics crystal ($\lambda=0$),
while the green-dashed ($\lambda=g$) and red-dotted curves ($\lambda=1.1 g$) correspond to that of 2D optomechanics.
The other parameters are: $\omega_{b}/(2\pi)=\omega_{c}/(2\pi)=100$ MHz,
$g/(2\pi)=20$ MHz, $|\Omega|/(2\pi)=0.02$ MHz, $\kappa_{a}/(2\pi)=1$ MHz,
$\kappa_{b}/(2\pi)=0.1$ MHz, $\kappa_{c}/(2\pi)=1000$ Hz, and $T=0.01$ K.
  Inset figure: steady photon numbers.}
\label{fig3}
\end{figure}

The numerical calculation results of the photon second-order correlation
 function in degenerate 2D optomechanical system are shown in Fig.~\ref{fig3}.
The blue-solid curve describes photon blockade of standard optomechanics, which is equivalent to a large gap between graphene and photonic crystal cavity so that their interaction can be neglected. The peaks and dips in blue-solid curve can be explained analytically as follows. In weak pumping limits $g\sqrt{n_a}|\Omega|/\omega_{b}\ll \omega_{b}$ and $\lambda\sqrt{n_a}|\Omega|/\omega_{c}\ll \omega_{c}$, the Hamiltonian $H_{r}$ in Eq.~(\ref{eq:3}), experiencing an unitary transformation $U_{1}$, becomes $H_{wl}\approx-\hbar(\Delta+\chi_{t})\hat{a}^{\dagger}\hat{a}-\hbar\chi_{t} \hat{a}^{\dagger}\hat{a}^{\dagger}\hat{a}\hat{a}+\hbar\omega_{b}\hat{b}^{\dagger}\hat{b}+\hbar\omega_{c}\hat{c}^{\dagger}\hat{c}+i(\Omega \hat{a}^{\dagger}-\Omega^{\ast}\hat{a})$.
 There is no direct interaction between photon and phonons in $H_{wl}$, if we neglect the excitations of phonons \cite{Li,Bamba,Savona,xu1}; thus, the low excited states wave-function of 2D optomechanics can be assumed as
 $|\psi\rangle=(A_{000}|0\rangle+A_{100}|1\rangle+A_{200}|2\rangle)\otimes |0_{b}\rangle \otimes |0_{c}\rangle$, then the second-order correlation function could be obtained as:
\begin{eqnarray}\label{eq:5}
g^{(2)}(0)\approx\frac{2|A_{200}|^{2}}{|A_{100}|^{4}}\approx4\left|\frac{\gamma_{a}+i2(\Delta-\chi_{t})}{\gamma_{a}+i(\Delta-2\chi_{t})}\right|^{2}.
\end{eqnarray}
 Equation~(\ref{eq:5}) could be used to explain single-photon blockade dip and two-photon resonant peak in Fig.~\ref{fig3}. The dip at $\Delta=-\chi_{b}$ in blue-solid curve satisfies $g^{(2)}(0)<1$ and describes single photon blockade of standard optomechanics($\epsilon=0$), and the peak at $\Delta=-2\chi_{b}$ satisfies $g^{(2)}(0)>1$ and corresponds to two-photon resonant transition.  The small peak at $\Delta=-3\chi_{b}$ corresponds to photon resonant transitions between states $|1,\tilde{n}_{b},\tilde{n}_{c}\rangle$ and $|2,\tilde{n}_{b},\tilde{n}_{c}\rangle$ \cite{xu1}.

The green-dashed and red-dotted curves in Fig.~\ref{fig3} describe  photon blockade in 2D optomechanics ($\lambda\neq 0$), which is equivalent to a limited gap between the graphene sheet and photonic crystal cavity.
  In the case of $\lambda=g$ and $\omega_{g}=\omega_{b}$, the photon nonlinearity coefficient in green-dashed curve is doubled compared with standard optomechanics (blue-solid curve), that is $\chi_{t}=2\chi_{b}$. So the position of single photon blockade dip (two-photon resonant peak) shifts to $\Delta=-2\chi_{b}$ ($\Delta=-4\chi_{b}$) in green-dashed curve.

 The coupling strength $\lambda=1.1g$ in the red-dotted curve, the positions of photon blockade dip or multi-photon resonant transition peaks in $x$-axis deviate from those of green-dashed curve ($\lambda=g$). From above discussions,  it is shown that the graphene resonator can be used to tune photon nonlinearity, photon blockade, and photon tunneling in 2D optomechanical system.

\subsection{Optomechanically induced transparency}

\begin{figure}
\includegraphics[bb=10 70 590 485, width=8.0 cm, clip]{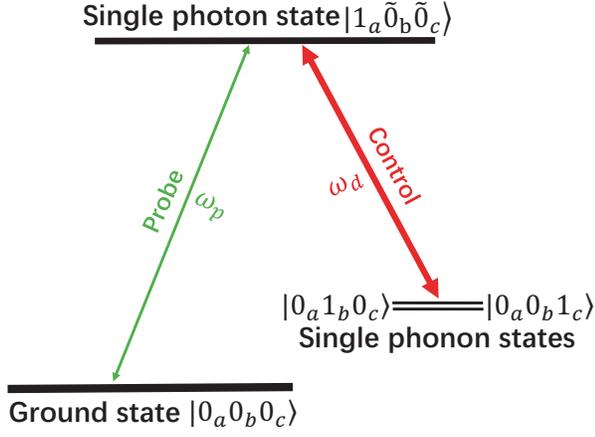}
\caption{(Color online) Schematic diagram for the 2D optomechanical system
  with the driving  (control)  and probe  fields with a single-particle excitation.
The vibration frequencies of two mechanical resonators are approximate the same ($\omega_{b}\approx\omega_{c}$),
the single excitation states form a $\Lambda$-type degenerate four-level system, where $|0_{a}1_{b}0_{c}\rangle$
and $|0_{a}0_{b}1_{c}\rangle$ are single phonon excited states.}
\label{fig4}
\end{figure}

The OMIT in optomechanical system has been theoretically and experimentally studied \cite{Agarwal,Kippenberg,Painter,Luki}.
 The mechanical resonator, TLS, and condensed states coupling to optical or mechanical mode of optomechanical system can affect photon transmission in optomechanical system \cite{Luki,Ma,Bienert,Hui1,Hui2,Hui3}.
If the frequencies of nanobeam and graphene resonators are approximately the same ($\omega_{b}\approx\omega_{c}$), the ground state and single excited states in 2D optomechanical system form a degenerate four-level structure as shown in Fig.~\ref{fig4}. Thus, some additional photon transition channels $|1_{a}\tilde{0}_{b}\tilde{0}_{c}\rangle\leftrightarrow |0_{a}0_{b}1_{c}\rangle$  are created by graphene resonator, this should affect photon transition of 1D optomechanical crystal.  The energy level structure in Fig.\ref{fig4} is similar to that of the hybrid TLS-optomechanics system, where the TLS splits phonon energy levels and leads to double-transparency windows \cite{Hui1}.

If the photonic crystal cavity is driven by  a  strong driving field and a weak probe field, that is $H_{drive}=i\hbar[\Omega\exp{(-i\omega_{d}t)}+\varepsilon \exp{(-i\omega_{p}t)} ] \hat{a}^{\dagger}+h.c.$, where $\Omega$ (or $\omega_{d}$) and $\varepsilon$  (or $\omega_{p}$) are the amplitudes (or frequencies) of the driving and probe fields, respectively.
  In a rotating frame at the driving field frequency $\omega_{d}$, with the mean field approximation,
   the steady value equations for photonic and mechanical modes are obtained as
\begin{eqnarray}\label{eq:6}
\langle\dot{\hat{a}}\rangle&=& \left(i\Delta-\frac{\kappa_{a}}{2}\right)\langle \hat{a}\rangle+\Omega+\varepsilon\exp{(-i\Delta_{p}t)}\nonumber\\
& &-ig\langle \hat{a}\rangle \left(\langle \hat{b}^{\dagger}\rangle+\langle \hat{b}\rangle\right)-i\lambda \langle \hat{a}\rangle\left(\langle \hat{c}^{\dagger}\rangle+\langle \hat{c}\rangle\right),\nonumber\\
\langle \dot{\hat{b}}\rangle&=& -\left(i\omega_{b}+\frac{\kappa_{b}}{2}\right)\langle \hat{b}\rangle-ig\langle \hat{a}^{\dagger}\rangle\langle \hat{a}\rangle,\\
\langle\dot{\hat{c}}\rangle&=& -\left(i\omega_{c}+\frac{\kappa_{c}}{2}\right)\langle \hat{c}\rangle-i\lambda \langle \hat{a}^{\dagger}\rangle \langle \hat{a}\rangle\nonumber.
\end{eqnarray}
where $\Delta_{p}=\omega_{p}-\omega_{d}$ is the frequency detuning between probe and driving fields. When the amplitude of pumping field is much larger than that of the probe field ($|\Omega| \gg|\varepsilon|$),  up to the first order small quantity of $\varepsilon$, the solutions of Eqs.~(\ref{eq:6}) can be approximately expanded as $\langle \hat{o}\rangle=O_{0}+ O_{+}\exp{(i\Delta_{p}t)}+O_{-}\exp{(-i\Delta_{p}t)}$,
with $o=a,b,c$ and $O=A,B,C$. Here $O_{0}$ and $O_{\pm}$ are respectively the steady values and first-order responses to weak probe field, and $|O_{0}|\gg |O_{\pm}|$.
Then the steady value of cavity field can be calculated as $A_{0}=\Omega/[ \kappa_{a}/2-i\Delta+2ig\Re(B_{0})
+2i\lambda \Re(C_{0})]$, and mechanical modes' steady values are $B_{0}=-ig|A_{0}|^{2}/(i\omega_{b}+\kappa_{b}/2)$ and $C_{0}=-i\lambda|A_{0}|^{2}/(i\omega_{c}+\kappa_{c}/2)$, respectively.

The  first-order small quantity $A_{-}$ describes probe field's absorption and dispersion in 2D optomechanical system, and it could be calculated from Eqs.~(\ref{eq:6}) as
\begin{equation}\label{eq:7}
A_{-}=\frac{\varepsilon}{Q-i\Delta_{p}-2igP|A_{0}|^{2}-\frac{4g^{2}P^{2}|A_{0}|^{4}}{Q-i\Delta_{p}+2igP|A_{0}|^{2}}}.
\end{equation}
with  $P=\sum_{i=b,c}{\theta_{i}\omega_{i}/[(\kappa_{i}/2-i\Delta_{p})^{2}+\omega^{2}_{i}]}$ ($\theta_{b,c}=g,\lambda$),
and $Q=\kappa_{a}/2-i\Delta+2ig\Re(B_{0})+2i\lambda \Re(C_{0})$.
With the input-output relation, the output optical field can be written as
 $\varepsilon_{out}=(2\kappa_{a}A_{0}-\Omega)+(2\kappa_{a}A_{-}-\varepsilon) \exp{(-i\Delta_{p}t)} +2\kappa_{a}A_{+}\exp{(i\Delta_{p}t)}$ \cite{Agarwal,Zoller}.
Define the quadratures of field as  $\varepsilon_{T}=\mu_{p}+i\nu_{p}$, here $\mu_{p}=\kappa_{a}(A^{\ast}_{-}+A_{-})/\varepsilon$ describes probe field's absorption in 2D optomechanical system, while the imaginary part $\nu_{p}=\kappa_{a}(A^{\ast}_{-}-A_{-})/\varepsilon$ corresponds to dispersion \cite{Agarwal}.

\begin{figure}
\includegraphics[bb=10 5 550 435, width=4.25 cm, clip]{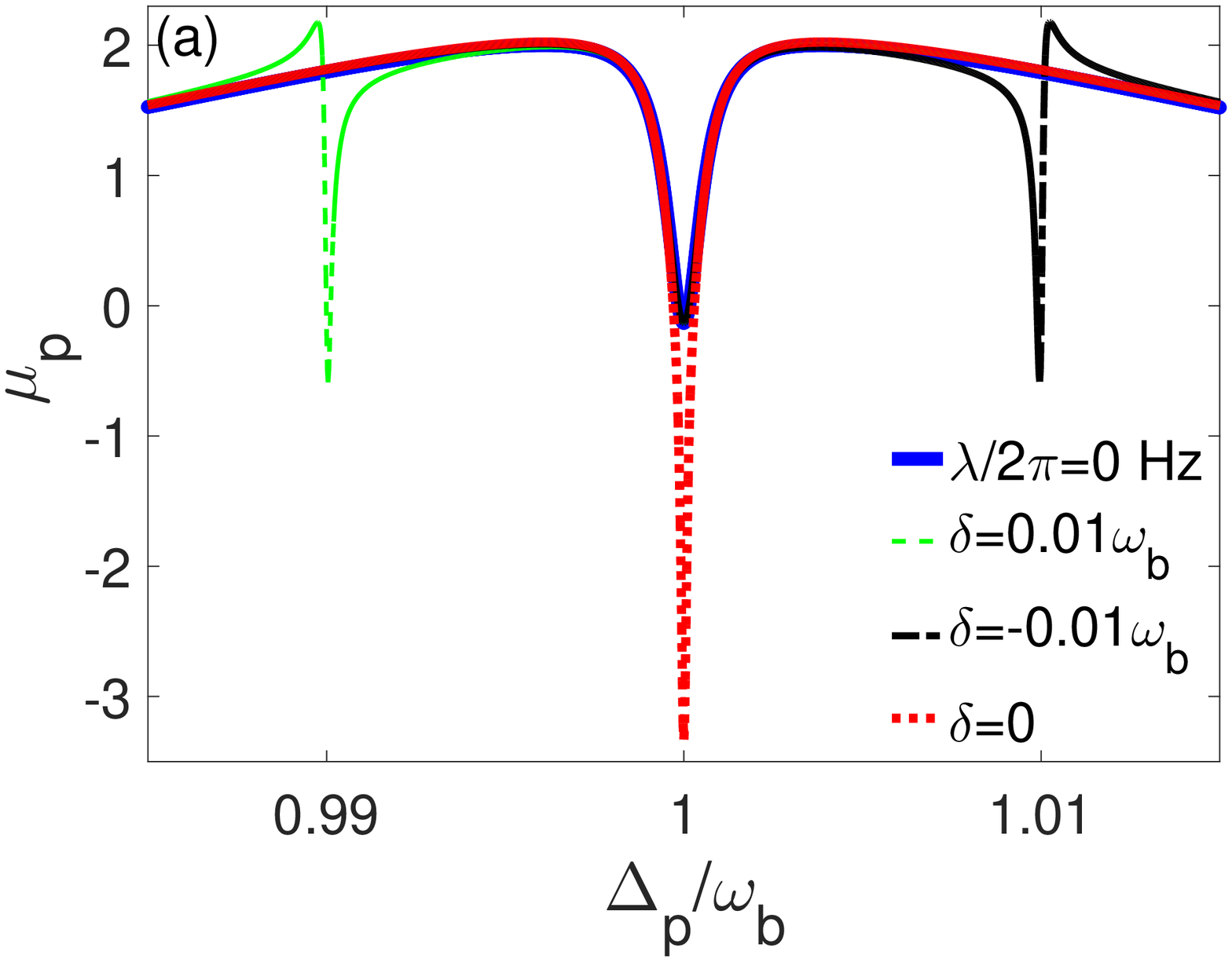}
\includegraphics[bb=12 0 560 440,  width=4.25 cm, clip]{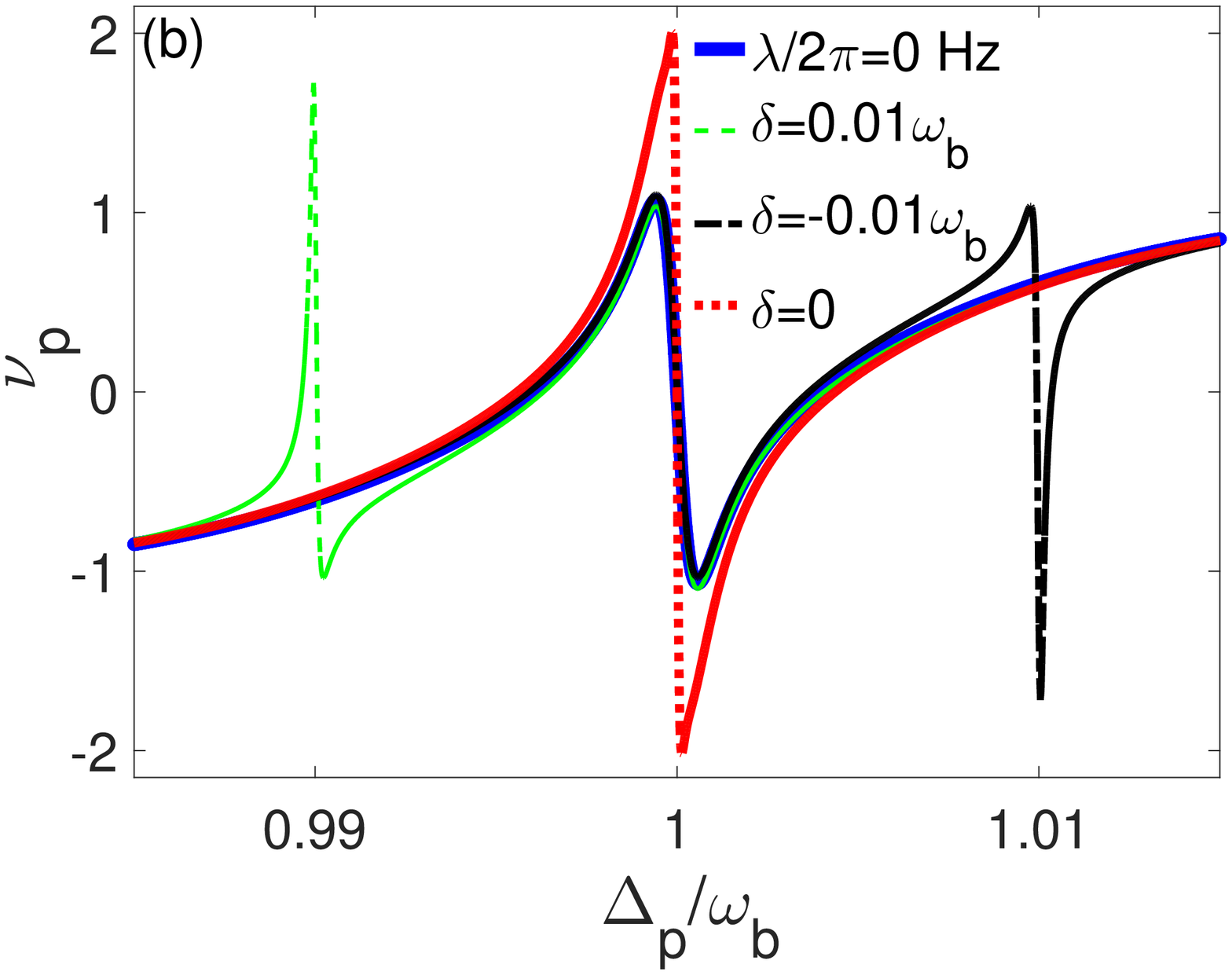}\\
\includegraphics[bb=10 5 580 430, width=4.27 cm, clip]{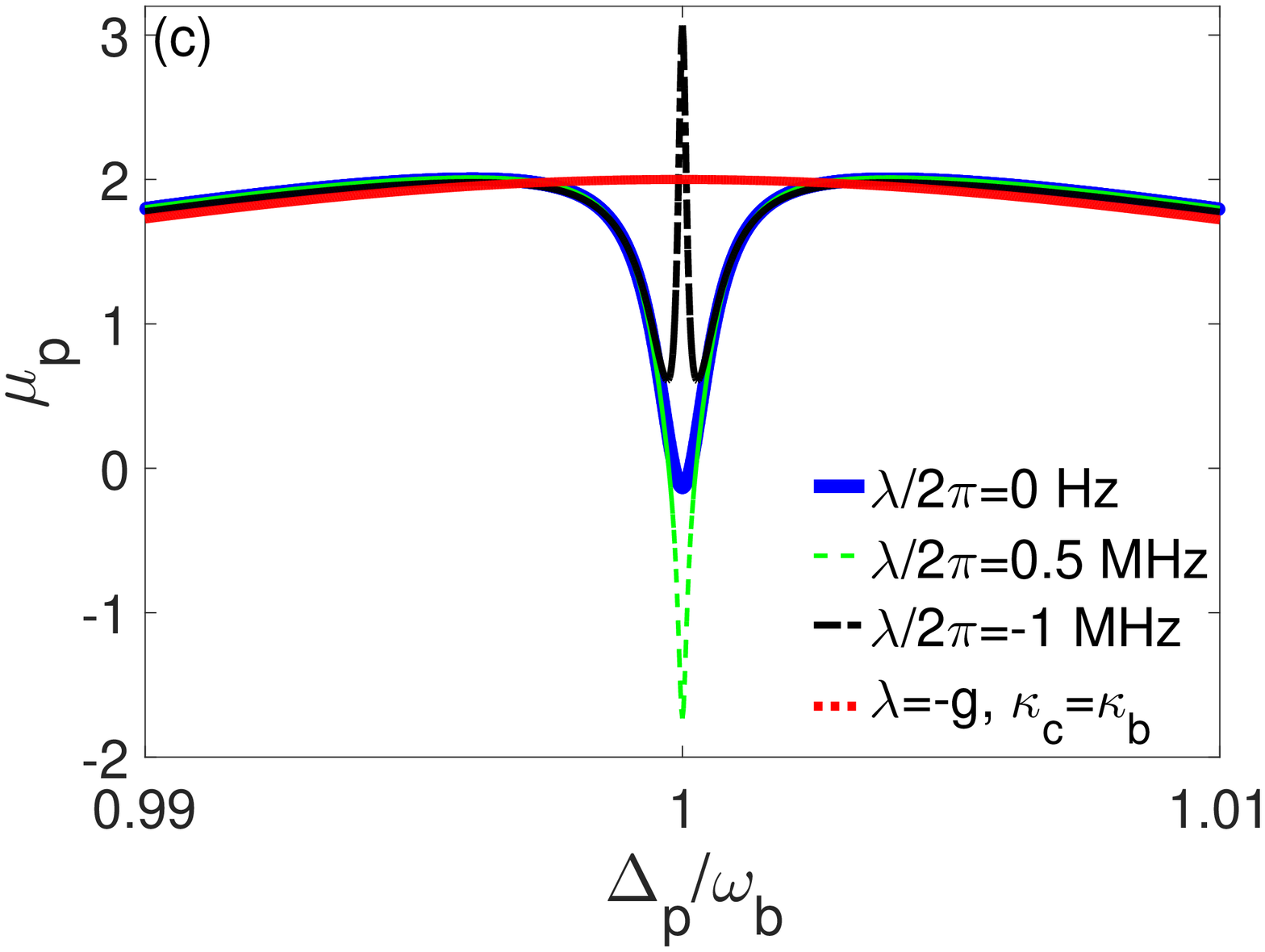}
\includegraphics[bb=10 0 575 440, width=4.27 cm, clip]{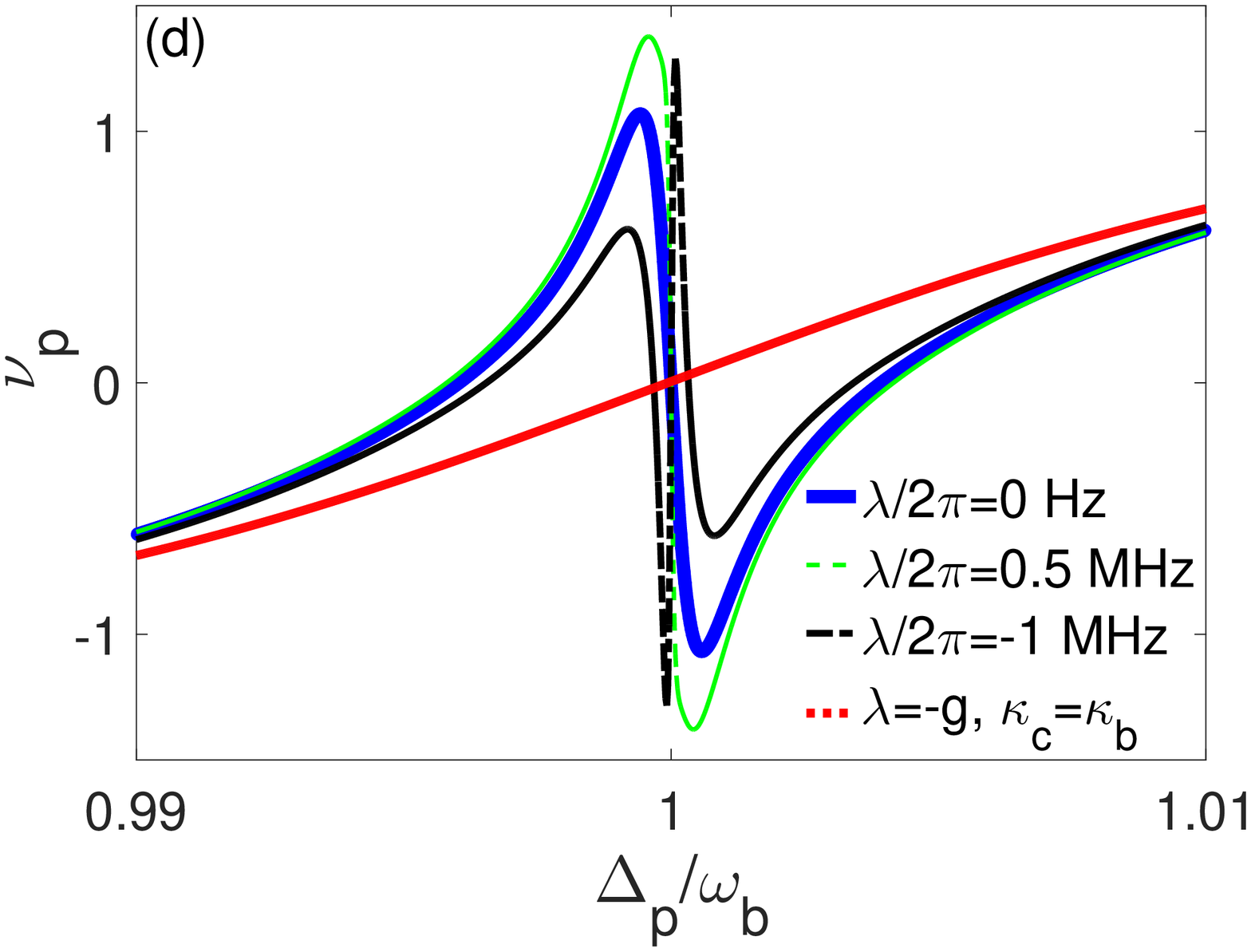}
\caption{(Color online) Photon transmission in 2D optomechanical system.
Probe field's  (a) absorption and (b) dispersion  for different $\delta =\omega_{b}-\omega_{c}$:
(i) $\lambda/(2\pi)=0$ Hz  (blue-solid); (ii) $\delta=0.01\omega_{b}$ (green-dashed); (iii) $\delta=-0.01\omega_{b}$ (black-dashed-dotted); 4) $\delta=0$ (red-dotted). We specify $\lambda/(2\pi)=1$ MHz and $\kappa_{c}/(2\pi)=0.01$ MHz (except the blue-solid curve).
For the same mechanical frequencies ($\omega_{b}=\omega_{c}$), the probe field's (c) absorption  and (d) dispersion for different $\lambda$: (i) $\lambda/(2\pi)=0$ Hz (blue-solid); (ii) $\lambda/(2\pi)=0.5$ MHz (green-dashed); (iii) $\lambda/(2\pi)=-1$ MHz (black-dashed-dotted); (iv) $\lambda=-g$ (red-dotted). The mechanical damping rates are the same $\kappa_{c}=\kappa_{b}$ in the red-dotted curve , but for other three curves $\kappa_{c}/(2\pi)=0.01$ MHz.
The other parameters of four figures are: $\omega_{b}/(2\pi)=100$ MHz, $\Delta=-\omega_{b}$, $\kappa_{a}/(2\pi)=5$ MHz,
$\kappa_{b}/(2\pi)=0.06$ MHz,  $g/(2\pi)=4$ MHz, and $|\Omega|/(2\pi)=10$ MHz.}
\label{fig5}
\end{figure}

 The effects of mechanical resonators' detuning $\delta=\omega_{b}-\omega_{c}$ on transparency window are shown in Figs.~\ref{fig5}(a) and \ref{fig5}(b).  There is no graphene-cavity interaction in blue-solid curve which describes the OMIT of standard optomechanics. When the frequencies of two mechanical resonators are the same ($\delta=0$),  the local minimal value within the transparency window moves down to negative regime in red-dotted curve, this means that the graphene resonator leads to the amplification of weak probe field which should be transparent in the 1D optomechanical crystal (blue-solid curve). This phenomenon originates from the additional photon transition channels ($|1_{a}\tilde{0}_{b}\tilde{0}_{c}\rangle\leftrightarrow |0_{a}0_{b}1_{c}\rangle$)  induced by graphene resonator(see Fig.~\ref{fig4}), which changes the photon numbers of states $|1_{a}\tilde{0}_{b}\tilde{0}_{c}\rangle$ and also the destructive coherent transitions in the $\Lambda$ type three-level structure of 1D optomechanical crystal.  For a small mechanical detuning  $\delta=0.01\omega_{b}$ ($-0.01\omega_{b}$),  two transparency windows appear in the green-dashed (black-dashed-dotted) curve, the transparency window at $\Delta_{p}=\omega_{b}$ is the same as that of standard optomechanics (blue-solid curve) and a new narrow transparency window appears at $\Delta_{p}=\omega_{b}-\delta$. This is similar to the double-color transparency window in the TLS-optomechanics coupled systems \cite{Hui1}.

For the same mechanical frequencies ($\omega_{b}=\omega_{c}$), the effects of graphene-cavity optomechanical coupling strength $\lambda$ on transparency window are shown in Figs.~\ref{fig5}(c) and \ref{fig5}(d). If the motions of the nanobeam and graphene resonators are inphase ($\lambda=g$), the probe signal is amplified in the green-dashed curve. If motions are out of phase $\lambda=-g$, the black-dashed-dotted curve shows that a absorption peak appears within the transparency window of standard optomechanics (blue-solid curve). When the graphene and nanobeam resonators have the same frequencies, damping rates, and optomechanical coupling strengths, but oppositive motion phases ($\lambda=-g$), the transparency window vanishes in red-dot curve in  Fig.~\ref{fig5}(c). The destructive coherent transitions induced by two mechanical resonators offset with each other. Through above discussions, it is shown that the graphene resonator can  control photon transmission in one-dimensional optomechanics crystal.

\section{ Ground state cooling }

\subsection{Mechanical Spectrum}

In the vacuum weak coupling regime $g,\lambda\ll \kappa_{a}$, the photon operator can be split into $\hat{a}=\alpha+\hat{d}$, with an average coherent amplitude $\alpha$ and a small fluctuation term $\hat{d}$ \cite{Marquardt2,Painter3}.
From Eq.~(\ref{eq:3}), we can obtain  the linearized Heisenberg equations for $\hat{d}$,  $\hat{b}$, and  $\hat{c}$ as follows:
\begin{eqnarray}\label{eq:8}
\dot{\alpha}&=& \left(i\Delta-\frac{\kappa_{a}}{2}\right)\alpha+\Omega,\nonumber\\
\dot{\hat{d}}&=& \left(i\Delta-\frac{\kappa_{a}}{2}\right)\hat{d}-iG(\hat{b}^{\dagger}+\hat{b})-iK(\hat{c}^{\dagger}+\hat{c})\nonumber\\
& &+\sqrt{\kappa_{a}}\hat{a}_{in}(t),\\
\dot{\hat{b}}&=& -\left(i\omega_{b}+\frac{\kappa_{b}}{2}\right)\hat{b}-i(G^{\ast}\hat{d}+G \hat{d}^{\dagger})+\sqrt{\kappa_{b}}\hat{b}_{in}(t),\nonumber\\
\dot{\hat{c}}&=& -\left(i\omega_{c}+\frac{\kappa_{c}}{2}\right)\hat{c}-i(K^{\ast}\hat{d}+K \hat{d}^{\dagger})+\sqrt{\kappa_{b}}\hat{c}_{in}(t).\nonumber
\end{eqnarray}
with $G=g\alpha$ and $K=\lambda\alpha$.  Under strong pumping field, the classical and nonlinear terms $-ig\hat{d}(\hat{b}^{\dagger}+\hat{b})$, $-i\lambda \hat{d}(\hat{c}^{\dagger}+\hat{c})$, and $-ig\hat{d}^{\dagger}\hat{d}$  have been  neglected.
Thus, the linearized Hamiltonian of 2D optomechanics could be written as
 $H_{L}=-\hbar\Delta \hat{d}^{\dagger}\hat{d}+\hbar\omega_{b}\hat{b}^{\dagger}\hat{b}+\hbar\omega_{c}\hat{c}^{\dagger}\hat{c}
+\hbar(G \hat{d}^{\dagger}+G^{\ast}\hat{d}) (\hat{b}^{\dagger}+\hat{b})+\hbar (K\hat{d}^{\dagger}+K^{\ast}\hat{d})(\hat{c}^{\dagger}+\hat{c})$.
Here $\hat{o}_{in}(t)$($o=a,b,c$) are input noise operators of optical and mechanical modes, respectively;  their  environmental average values are $\langle \hat{o}_{in}(t)\rangle=0$, and the nonzero noise input correlation functions are $\langle \hat{o}_{in}(t^{\prime})\hat{o}^{\dagger}_{in}(t)\rangle=(n^{T}_{o}+1)\delta(t^{\prime}-t)$ and $\langle \hat{o}^{\dagger}_{in}(t^{\prime})\hat{o}_{in}(t)\rangle=n^{T}_{o}\delta(t^{\prime}-t)$.

With the Fourier transformation, the motion equations in frequency domain are as follows:
\begin{eqnarray}\label{eq:9}
-i\omega \tilde{d}(\omega)&=& \left(i\Delta-\frac{\kappa_{a}}{2}\right)\tilde{d}(\omega)-iG\left[\tilde{b}^{\dagger}(\omega)+\tilde{b}(\omega)\right]\nonumber\\
& &-iK\left[\tilde{c}^{\dagger}(\omega)+\tilde{c}(\omega)\right]+\sqrt{\kappa_{a}}\tilde{a}_{in}(\omega),\nonumber\\
-i\omega \tilde{b}(\omega)&=& -\left(i\omega_{b}+\frac{\kappa_{b}}{2}\right)\tilde{b}(\omega)+\sqrt{\kappa_{b}}\tilde{b}_{in}(\omega)\nonumber\\
& &-i\left[G^{\ast}\tilde{d}(\omega)+G \tilde{d}^{\dagger}(\omega)\right],\\
-i\omega\tilde{c}(\omega)&=& -\left(i\omega_{c}+\frac{\kappa_{c}}{2}\right)\tilde{c}(\omega)+\sqrt{\kappa_{c}}\tilde{c}_{in}(\omega)\nonumber\\
& &-i \left[K^{\ast}\tilde{d}(\omega)+K \tilde{d}^{\dagger}(\omega)\right].\nonumber
\end{eqnarray}
 Eliminating the variables $\tilde{d}^{\dagger}(\omega)$, $\tilde{b}(\omega)$, $\tilde{b}^{\dagger}(\omega)$, $\tilde{c}(\omega)$, and $\tilde{c}^{\dagger}(\omega)$ in Eqs.~(\ref{eq:9}), with $\tilde{o}^{\dagger}(\omega)=[\tilde{o}(-\omega)]^{\ast}$ ($o=b,c,d$), then $\tilde{b}(\omega)$ can be expressed with input noise operators $\tilde{a}_{in}(\omega)$, $\tilde{b}_{in}(\omega)$, and $\tilde{c}_{in}(\omega)$ \cite{Mancini}.
The mechanical spectrum  of nanobeam resonator: $S_{bb}(\omega)=\int^{+\infty}_{-\infty}{\langle\tilde{b}^{\dagger}(\omega^{\prime})\tilde{b}(\omega)\rangle d\omega^{\prime}}$ \cite{Marquardt2,Painter3}:
\begin{equation}\label{eq:10}
S_{bb}(\omega)= \frac{\frac{|\alpha|^{2}}{\kappa_{a}}\sigma_{opt}(\omega)+\kappa_{b}\sigma^{(b)}_{th}(\omega)+\kappa_{c}\sigma^{(c)}_{th}(\omega)}{|N(\omega) |^{2}}.
\end{equation}
where
\begin{eqnarray}\label{eq:11}
\sigma_{opt}&=&\kappa^{2}_{a}g^{2}|\Gamma^{-1}_{b}(\omega)|^{2} |\Gamma_{a}(\omega)|^{2},\nonumber\\
\sigma^{(b)}_{th}&=&n^{T}_{b} |\Gamma^{-1}_{b}(\omega)\Sigma_{c}(\omega)+i\Sigma(\omega)|^{2}+(n^{T}_{b}+1) |\Sigma(\omega)|^{2},\nonumber\\
\sigma^{(c)}_{th}&=&\epsilon^{2} |\Gamma^{-1}_{b}(\omega)|^{2} |\Sigma(\omega)|^{2}\nonumber\\
& &\times\left[(n^{T}_{c}+1)|\Gamma_{c}(\omega)|^{2}+n^{T}_{c}|\Gamma_{c}(-\omega)|^{2}\right],\\
N&=&\Gamma^{-1}_{b}(\omega)[\Gamma^{-1}_{b}(-\omega)]^{\ast}\Sigma_{c}(\omega)+2\omega_{b}\Sigma(\omega).\nonumber
\end{eqnarray}
with $\epsilon=K/G=K^{\ast}/G^{\ast}=\lambda/g$, with $g$ and $\lambda$ are assumed as real numbers.  The optical and mechanical responsive functions are defined as $\Gamma_{a} (\omega)=[\kappa_{a}/2-i(\omega+\Delta)]^{-1}$,  $\Gamma_{b}(\omega)=[\kappa_{b}/2+i(\omega_{b}-\omega)]^{-1}$,
and $\Gamma_{c}(\omega)=[\kappa_{c}/2+i(\omega_{c}-\omega)]^{-1}$, respectively. The self-energy term of optomechanical crystal is defined as $\Sigma(\omega)=-i|G|^{2}\left[\Gamma^{-1}_{a} (\omega)-[\Gamma^{-1}_{a}(-\omega)]^{\ast}\right]$, which decides  frequency shift and extra damping rate of nanobeam resonator in standard optomechanics.

\begin{figure}
\includegraphics[bb=0 30 590 527, width=8.6 cm, clip]{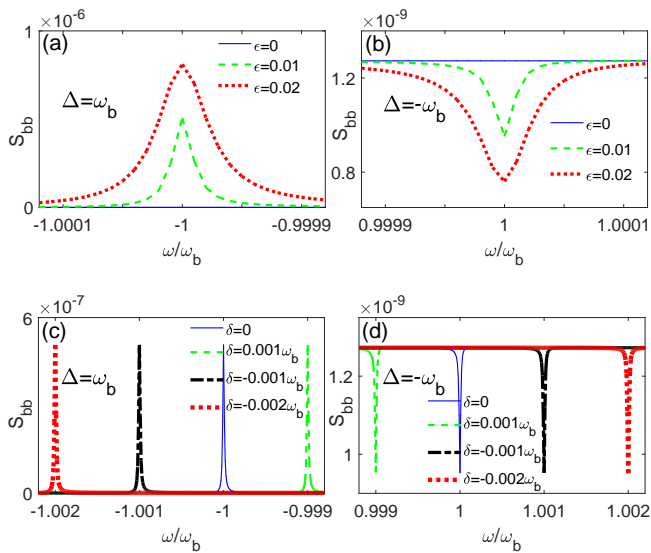}
\caption{(Color online) Tunable Mechanical spectrum.
 The mechanical spectrum under (a) blue-detuning laser   and (b) red-detuning laser for different coupling strength $\lambda$: (i) $\epsilon=0$ (blue-solid);  (ii)  $\epsilon=0.01$ (green-dashed); (iii) $\epsilon=0.02$ (red-dotted), here $\delta=0$.
The mechanical spectrum under blue-detuning laser (c) and red-detuning laser (d) for different detuning $\delta$: (i) $\delta=0$ (blue-solid);  (ii) $\delta=0.001\omega_{b}$ (green-dashed);  (iii) $\delta=-0.001\omega_{b}$  (black-dashed-dotted); (iv) $\delta=-0.002\omega_{b}$  (red-dotted), here $\epsilon=0.01$.
The other parameters of four figures are: $\omega_{b}/(2\pi)=100$ MHz, $\kappa_{a}/(2\pi)=10$ MHz, $\kappa_{b}/(2\pi)=10$ MHz, $\kappa_{c}/(2\pi)=1000$ Hz, $|G|/(2\pi)=1.3$ GHz, $n^{T}_{a}=0$, $n^{T}_{b}=100$, and $n^{T}_{c}=100$.}
\label{fig6}
\end{figure}

 Here $\sigma_{opt}$ describes the contribution of the optomechanical interaction between the nanobeam resonator and cavity field, $\sigma^{(b)}_{th}$ originates from the nanobeam resonator's thermal noises, and $\sigma^{(c)}_{th}$ comes from graphene resonator. Compared with standard optomechanics, an effective amplification factor  $\Sigma_{c}(\omega)=1+2\omega_{c} \epsilon^{2}\Sigma(\omega)/[(\kappa_{c}/2-i\omega)^{2}+\omega^{2}_{c}]$, which is induced by the graphene resonator, appears in nanobeam resonator's responsive function (see Eq.~(\ref{eq:11})). The equations $[\Sigma(-\omega)]^{\ast}=\Sigma(\omega)$ and
$[\Sigma_{c}(-\omega)]^{\ast}=\Sigma_{c}(\omega)$ have been adopted during the above calculations.

The blue-solid curve in Fig.~\ref{fig6}(a) describes mechanical spectrum of standard optomechanics($\lambda/(2\pi)=0$ Hz), which has been widely studied in both theories and experiments \cite{Vahala, Aspelmeyer1,Marquardt2}. If $\epsilon\neq0$,
under blue-detuning driving laser ($\Delta=\omega_{b}$), the narrow peaks appear at $\omega=-\omega_{b}$ in green-dashed and red-dotted curves representing the local amplification for mechanical spectrum. The FWHM (Full width at half maximum width) of narrow peak is close to $\kappa_{c}$ ($\kappa_{c}\ll\kappa_{b}$),
which indicates that the narrow peak (or dip) originates from the graphene resonator. For the red-detuning driving laser ($\Delta=-\omega_{b}$) in Fig.~\ref{fig6}(b), a narrow dip appears at $\omega=\omega_{b}$ in green-dashed and red-dotted  curves, which corresponds to local  suppression for mechanical spectrum. The dip in red-dotted curve is deeper than that of green-dashed curve, which means that a larger suppression for  mechanical spectrum of nanobeam resonator could be realized with a larger graphene-cavity interaction strength. The numerical calculations show that the amplification factor $\Sigma_{c}(\omega)$ is responsible for the narrow peak or dip in the mechanical spectrum of the graphene resonator. For a weak graphene-cavity coupling strength, $\Sigma_{c}(\omega)\approx 1$, the narrow peak or dip will disappear.

 The effects of mechanical detuning $\delta=\omega_{b}-\omega_{c}$ on mechanical spectrum of nanobeam resonator are shown in Fig.~\ref{fig6}(c) (blue-detuning) and \ref{fig6}(d) (red-detuning). For nonzero mechanical detuning ($\delta\neq0$), under blue detuning driving laser, the $x$-axis positions of narrow peaks in Fig.~\ref{fig6}(c) shift to $\omega=\omega_{b}+\delta$.
For red detuning driving laser in Fig.~\ref{fig6}(d), the narrow dips  shift to $\omega=\omega_{b}-\delta$. Thus, we can control the positions of amplification peaks and suppression dips in  mechanical spectrum of nanobeam resonator by tuning the frequency of the graphene resonator.

\subsection{Dynamical Backaction}

The optomechanical interaction changes mechanical resonator's frequencies and damping rates \cite{Vahala,Painter2,Thompson,Aspelmeyer1,Marquardt2}, which can be used for mechanical self-oscillation, phonon laser, mechanical ground state cooling,  and so on \cite{Barton,Cohadon,Zhang,Capuj,Park,Durand,Grudinin,Yang}. In this section we discuss the shifts of nanobeam resonator's frequency and damping rate in 2D optomechanical system.

If the frequencies of graphene and nanobeam resonators are approximately the same ($\omega_{b}\approx\omega_{c}$),
we can obtain from Eqs.~(\ref{eq:11}) that $N(\omega)\approx\Gamma^{-1}_{b}(\omega) [\Gamma^{-1}_{b}(-\omega)]^{\ast}+2\omega_{b}\left[ \epsilon^{2}\eta(\omega)+1\right]\Sigma(\omega)$,
with $\eta(\omega)=\Gamma_{c}(\omega) [\Gamma_{c}(-\omega)]^{\ast}/ \{\Gamma_{b}(\omega) [\Gamma_{b}(-\omega)]^{\ast}\}$.
The effective optomechanical self-energy in 2D optomechanics can be defined as $\Sigma_{e}(\omega)=[ \epsilon^{2}\eta(\omega)+1]\Sigma(\omega)$. In weak coupling limit $\kappa_{b}, \kappa_{c}, \Gamma_{opt}\approx|\alpha|^{2}/\kappa_{a}\ll \kappa_{a}$, the frequency shift and extra damping rate of nanobeam resonator can be defined as $\delta\omega^{(o)}_{b}=\Re [\Sigma_{e}(\omega_{b})]$ and $\gamma^{(o)}_{b}=-2\Im[\Sigma_{e}(\omega_{b})]$, respectively.
  The graphene resonator's frequency shift  $\delta\omega^{(o)}_{c}$ and extra damping rate  $\gamma^{(o)}_{c}$  can be calculated similarly.  Because of the symmetry of two resonators, the $\delta\omega^{(o)}_{c}$  and $\gamma^{(o)}_{c}$ can be simply obtained from $\delta\omega^{(o)}_{b}$ and $\gamma^{(o)}_{b}$ by interchanging subscript $b$ and $c$ in corresponding expressions, respectively.

\begin{figure}
\includegraphics[bb=0 35 550 400, width=8.5 cm, clip]{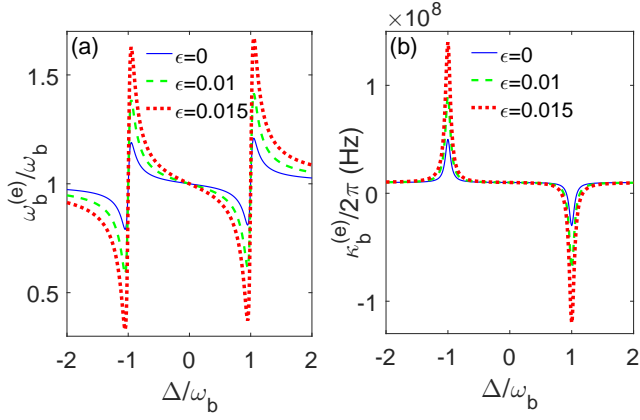}
\caption{(Color online) Nanobeam resonator's effective frequency  $\omega^{(e)}_{b}=\omega_{b}+\delta\omega^{(o)}_{b}$ and damping rate $\kappa^{(e)}_{b}=\kappa_{b}+\gamma^{(o)}_{b}$ in the resolved-sideband regime.
 The other parameters are:  $\omega_{b}/(2\pi)=\omega_{c}/(2\pi)=100$ MHz, $\kappa_{a}/(2\pi)=10$ MHz, $\kappa_{b}/(2\pi)=10$ MHz, $\kappa_{c}/(2\pi)=1000$ Hz, $|G|/(2\pi)=10$ MHz, $n^{T}_a=0$, $n^{T}_b=100$, and $n^{T}_c=100$.}
\label{fig7}
\end{figure}

The nanobeam resonator's effective frequency $\omega^{(e)}_{b}=\omega_{b}+\delta\omega^{(o)}_{b}$  and damping rate $\kappa^{(e)}_{b}=\kappa_{b}+\gamma^{(o)}_{b}$ in resolved-sideband regime ($\omega_{b},\omega_{c}\gg\kappa_{a}$) are shown in Figs.~\ref{fig7}(a) and \ref{fig7}(b), respectively.
The blue-solid curves in Figs.~\ref{fig7}(a) and \ref{fig7}(b) correspond to  effective frequency and damping rate of standard optomechanics ($\epsilon=0$).
With the increase of graphene-cavity optomechanical coupling strength, the changes of effective frequency and damping rate become larger in green-dashed and red-dotted curves, which means that graphene resonator can assist mechanical amplification and ground state cooling in 2D optomechanical system.

\begin{figure}
\includegraphics[bb=0 15 595 430, width=8.5 cm, clip]{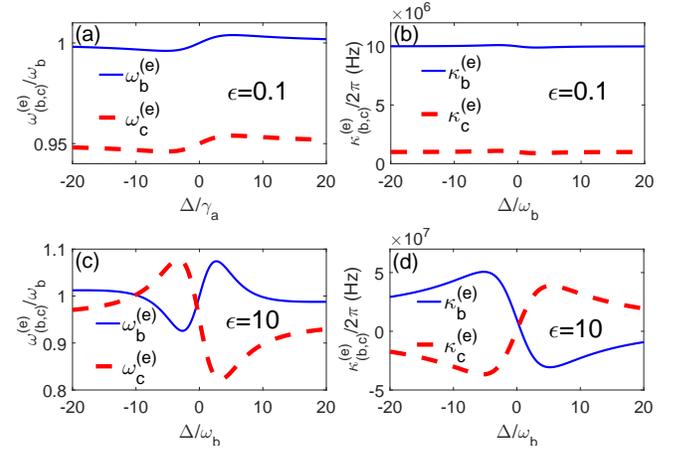}
\caption{(Color online) Coherent mechanical mixing in the unresolved-sideband regime. Here (a) and (c) describe effective frequencies ($\omega^{(e)}_{b,c}=\omega_{b,c}+\delta\omega^{(o)}_{b,c}$) of nanobeam (blue-solid) and graphene (red-dashed) resonators, while (b) and (d) correspond to their effective damping rates ($\kappa^{(e)}_{b,c}=\kappa_{b,c}+\gamma^{(o)}_{b,c}$ ). For (a) and (b) $\epsilon=0.1$, while $\epsilon=10$ for (c) and (d). The other parameters of four figures are:  $\omega_{b}/(2\pi)=100$ MHz,  $\omega_{c}/(2\pi)=95$ MHz, $\kappa_{a}/(2\pi)=1$ GHz, $\kappa_{b}/(2\pi)=10$ MHz, $\kappa_{c}/(2\pi)=1$ MHz,  $|G|/(2\pi)=10$ MHz, $n^{T}_a=0$, $n^{T}_b=100$, and $n^{T}_c=100$.}
\label{fig8}
\end{figure}

 According to recent experiments, the coupled mechanical system shows coherent mixing of mechanical modes in the bad-cavity limit\cite{Rosenberg,Zhou}. The changes of effective frequencies and damping rates for two mechanical resonators  in unresolved-sideband regime ($\omega_{b},\omega_{c}\ll \kappa_{a}$) are shown in Fig.~\ref{fig8}, where $\omega^{(e)}_{b,c}=\omega_{b,c}+\delta\omega^{(o)}_{b,c}$  and $\kappa^{(e)}_{b,c}=\kappa_{b,c}+\gamma^{(o)}_{b,c}$.
 If $\lambda\ll g$, the changes of two resonators' effective frequencies (Fig.~\ref{fig8}(a)) and damping rates (Fig.~\ref{fig8}(b)) are always in step, the effects of one mechanical resonator on  another resonator's optomechanical dynamical backaction are very weak.
 For a large graphene-cavity coupling strength($\lambda\sim g$),  the shifts of effective frequencies (Fig.~\ref{fig8}(c)) and  damping rates (Fig.~\ref{fig8}(d)) in blue-solid and red-dashed curves are the opposite of paces.  The damping rates of nanobeam and graphene resonators always have oppositive signs in Fig.~\ref{fig8}(c), this means when one mechanical resonator is amplified, the other one should be simultaneously cooled, moreover similar results can also be obtained in resolved-sideband regime. The blue-solid and red-dashed curves in Figs.~\ref{fig8}(c) and \ref{fig8}(d) cross with each other, which means that coherent mixing of two mechanical resonators can be realized in 2D optomechanical system with optical field serving as an intermediacy.

\subsection{Graphene assisted Ground State Cooling}

As shown in Fig.~\ref{fig7}, the graphene resonator could enhance damping rate of the nanobeam resonator, which should make positive contributions to the ground state cooling of the nanobeam resonator. However, such a graphene resonator also introduces a noise term $\sigma^{(c)}_{th}$ in  Eq.~(\ref{eq:10})  which makes negative contributions to ground state cooling. The relative value of above two terms decide the net contribution of the graphene sheet to ground state cooling of the nanobeam resonator.
With the mechanical spectrum in Eq.~(\ref{eq:10}), the phonon number on nanobeam resonator can be obtained as \cite{Marquardt2,Painter3}
\begin{eqnarray}\label{eq:12}
 n_{m}= \int^{+\infty}_{-\infty}\frac{d\omega}{2\pi} S_{bb}(\omega)
\end{eqnarray}

\begin{figure}
\includegraphics[bb=0 10 580 520, width=8.6 cm, clip]{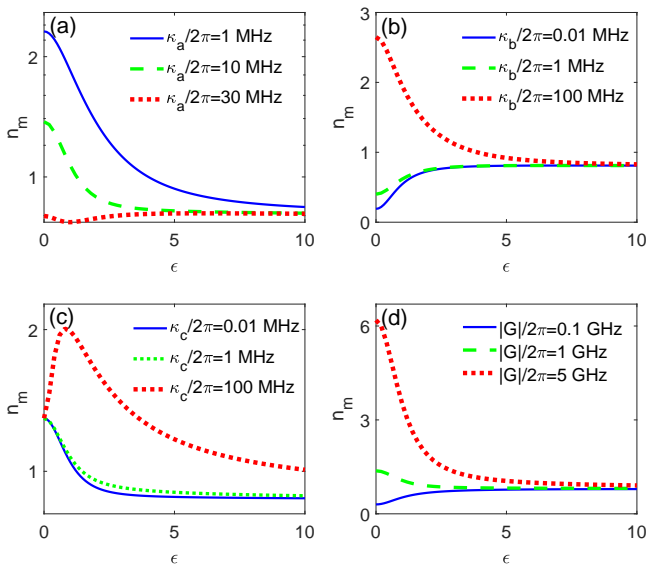}
\caption{(Color online) Graphene assisted Ground state cooling of the nanobeam resonator.
(a) The effects of cavity damping, with $\kappa_{b}/(2\pi)=10$ MHz, $\kappa_{c}/(2\pi)=0.1$ MHz, and $|G|/(2\pi)=1$ GHz;
(b) The effects of nanobeam damping, with $\kappa_{a}/(2\pi)=10$ MHz, $\kappa_{c}/(2\pi)=0.1$ MHz, and $|G|/(2\pi)=1$ GHz;
(c) The effects of graphene damping, with $\kappa_{a}/(2\pi)=10$ MHz, $\kappa_{b}/(2\pi)=10$ MHz, and $|G|/(2\pi)=1$ GHz;
(d) The effects of nanobeam-cavity coupling strength, with $\kappa_{a}/(2\pi)=10$ MHz, $\kappa_{b}/(2\pi)=10$ MHz, and $\kappa_{c}/(2\pi)=0.1$ MHz.
 The other parameters of four figures are: $\omega_{b}/(2\pi)=\omega_{c}/(2\pi)=100$ MHz, $\Delta=-\omega_{b}$, $n^{T}_a=0$, $n^{T}_b=5$, and $n^{T}_c=5$. }
\label{fig9}
\end{figure}

The variation of  phonon number on nanobeam resonator as a function of graphene-cavity coupling strength is shown in Fig.~\ref{fig9},  the value $n_{m}(\epsilon=0)$  corresponds to phonon number of standard optomechanics.
Figure 9(a) shows the effects of different cavity damping rates on the ground state cooling of nanobeam resonator.
In blue-solid (or green-dashed) curve of Fig.~\ref{fig9}(a), the phonon number of 1D optomechanical crystal is larger than $1$ ($n_{m}(\epsilon=0)>1$) for the case of no graphene-cavity interaction;
 however, the phonon number is quickly suppressed  and reaches to its ground state ($n_{m}< 1$) as the increase of graphene-cavity coupling strength. The  green-dashed and red-dotted curves show that the suppression ability of the graphene sheet on nanobeam resonator's phonon number  becomes weaker in the case of a larger cavity damping rate (compared with blue-solid curve).

 The ground state cooling of nanobeam resonators of different damping rates in 2D optomechanical system are shown in Fig.~\ref{fig9}(b).
 Compared with small damping rates of blue-solid and green-dashed curves, the phonon number on nanobeam resonator with larger damping rate  can be more effectively suppressed by the graphene resonator in red-dotted curve.  The effects of graphene resonator's damping rate and optomechanical coupling strength in 1D optomechanical crystal on the ground state cooling of nanobeam resonator are shown in  Figs.~\ref{fig9}(c) and \ref{fig9}(d), respectively.
From above results, it is clear that the graphene resonator can enhance ground state cooling of nanobeam resonator in some parameter regimes, especially for the high damping nanobeam resonator.

\section{ Discussion and Conclusion}

We proposed a new type of multi-mode optomechanics formed by a suspended graphene resonator coupled to the photonic mode of a 1D optomechanical crystal. This kind of 2D optomechanical system could be experimentally realized by transferring  graphene sheets above a released 1D photonic crystal cavity on silicon substrate, and the gap between graphene sheet and photonic cavity is the main difficulty for experimental success. By tuning the control voltage between graphene and silicon substrate, the steady-state position of the graphene sheet can be easily adjusted by a control voltage, and this can affect the damping rate of photonic cavity and graphene-cavity optomechanical interaction strength.

We have theoretically studied photon transmission and mechanical ground state cooling in the proposed 2D optomechanical system. Additional transitions channels induced by the graphene resonator could enhance or weaken the destructive coherent transitions of photons, and the tranmission of probe signal in 2D optomechanics can be switched among absorption, transparency, and amplification by tuning graphene sheet's vibration frequency and graphene-cavity optomechanical interaction strength.

 The graphene resonator could induce narrow amplification peak and suppression dip in the noise spectrum of  nanobeam resonator, and affects effective damping rate of the nanobeam resonator.  According to our numerical calculations,  the graphene resonator can assist the ground state cooling of nanobeam resonator in certain parameter regime.

 Additionally, the graphene sheet has some special properties, such as  nonlinear mechanical motion and  nonlinear  mechanical damping, which could  be interesting research topics in the future.

\section{ Acknowledgement}

Guangya Zhou is supported by MOE Academic Research Fund of Singapore under Grant No. R-265-000-557-112.  X.W.X. is supported by the National Natural Science Foundation of China (NSFC) under Grants No.11604096 and the Startup Foundation for Doctors of East China Jiaotong University under Grant No. 26541059.  Y.J.Z. is supported by the China Postdoctoral Science Foundation under grant No. 2017M620945.

\end{document}